# The Space Experiment of the Exo-ecosystem

Zhu Liu*, Duo Cui, Siyao Yang
Department of Earth System Science, Tsinghua University
*Email: zhuliu@tsinghua.edu.cn

**Abstract**

The experiment of exo-ecosystem and the exploration of extraterrestrial habitability aims to explore the adaptation of terrestrial life in space conditions for the manned space program and the future interstellar migration, which shows great scientific significance and public interests. By our knowledge the early life on Earth, archaea and extremophile have the ability to adapt to extreme environmental conditions and can potentially habitat in extraterrestrial environments. Here we proposed a design and framework for the experiment on exo-ecosystem and extraterrestrial habitability. The conceptual approach involves building an ecosystem based on archaea and extremophiles in a simulated extraterrestrial environment, with a focus on assessing the exobiological potential and adaptability of terrestrial life forms in such conditions through controlled experiments. Specifically, we introduce the Chinese Exo-Ecosystem Space Experiment (CHEESE), which investigates the survivability and potential for sustained growth, reproduction, and ecological interactions of methanogens under simulated Mars and Moon environments using the China Space Station (CSS) as a platform. We highlight that the space station provides unique yet relatively comprehensive conditions for simulating extraterrestrial environments. In conclusion, space experiments involving exo-ecosystems could pave the way for long-term human habitation in space, ensuring our ability to sustain colonies and settlements beyond Earth while minimizing our ecological impact on celestial bodies."

## 1. Introduction

The exploration and colonization of space have long been the aspirations of humanity. As we inch closer to turning these dreams into reality, it is essential to consider the construction of exo-ecosystems, self-sustaining habitats in extraterrestrial environments. "Exo-ecosystem" is taken to research the interaction of life and spacial environment, which includes the adaptation of earth life in extraterrestrial environments for future interstellar migrants, as well as exploring possible life or ecosystem forms that may exist in exoplanets. One of the primary motivations for constructing exo-ecosystems is the potential for groundbreaking scientific discoveries. Study life in extreme environments beyond Earth enable us to unravel life's adaptability and resilience, which could provide unique opportunities to explore the boundaries of life and gain insights into the origins of life. Furthermore, exo-ecosystems may offer us a glimpse into potential extraterrestrial biospheres, enabling us to identify and study ecosystems that could exist on other celestial bodies.

The experiment of the Exo-ecosystem aims to determine the survivability and potential for sustained growth, reproduction, and ecological interactions of Earth's organisms under simulated or actual extraterrestrial conditions. The investigation involves subjecting selected organisms to specific environmental parameters, such as temperature, atmospheric composition, radiation levels, gravity, and resource availability, relevant to the target extraterrestrial environment. Through rigorous observation, analysis, and comparison with known physiological and ecological characteristics, the study seeks to evaluate the compatibility and potential for establishing self-sustaining ecosystems or exo-ecosystems in extraterrestrial habitats.

The initial exo-ecosystem experiments were to verify the adaptation of Earth animals, plants, and microorganisms in simulated extraterrestrial environments constructed in the ground-based laboratory (Caporale et al., 2023; Kasiviswanathan et al., 2022; Wamelink et al., 2014). One-factor experiments are typically used to simulate extraterrestrial environments in ground-based laboratories. These experiments focus on isolating and studying the effects of individual factors, such as extremely low or high temperatures (Mickol et al., 2018), high pressures (Mickol and Kral, 2017), Martian soil analogs (Maus et al., 2020), and UV radiation (Sinha et al., 2017) on Earth's organisms. Nonetheless, the authentic extraterrestrial environment encompasses a multitude of conditions, rendering it significantly more intricate to replicate fully here on Earth. The progress in space stations, and Mars rovers has created an ideal platform to simulate extraterrestrial environments beyond Earth's confines. This has led to a gradual increase in space exo-ecosystem experiments, which are now being conducted on satellites, spacecraft, and the International Space Station (ISS). Previously, NASA, Russia, the European Space Agency ( ESA ), and other international partners conducted a series of studies on the International Space Station and on Earth (Dachev et al., 2017; Rabbow et al., 2017, 2015, 2012). Current advances demonstrate the progress being made in developing the necessary technologies and knowledge to construct and sustain exo-ecosystems. Continued research, experimentation, and interdisciplinary collaboration are key to furthering our understanding and capabilities in this exciting field of space exploration and colonization.

## 2. Previous space experiment of exo-ecosyetem

Extraterrestrial ecosystem experiments can be classified based on their location into three categories: near-space, satellites or spacecraft, and space stations. Each of these locations offers distinct extraterrestrial environmental conditions, providing unique advantages and opportunities for conducting simulation studies and enhancing our understanding of extraterrestrial ecosystems.

Earth's near space, situated between 20 and 100 km above sea level, stands as one of the unique analogues of both Mars' surface and the atmosphere of Venus, displaying conditions that include low atmospheric pressure, extremely low temperature and humidity, and a high flux of ultraviolet and cosmic radiation(Smith, 2013). To facilitate the study of exo-ecosystem experiments in near space, the Chinese Academy of Sciences has developed a balloon-borne astrobiology platform (CAS-BAP). This platform offers essential instruments for exposing biological samples to the environment, controlling temperature, collecting bioaerosols and dust in situ, and performing UV spectrometry(Lin et al., 2022; Wang et al., 2023). TC-BIOSEP had uploaded microbes, algae, and nematodes into the near-space region (Wang et al., 2023).

China's Jianbing series of satellites have deployed three sealed modules named CBS-1, CBS-2, and DM-11. Among these, CBS-1 accommodated two mice with provisions for oxygen supply, food supply, waste collection, and circadian rhythm monitoring, observing the mice's activities continuously(Li et al., 2021). These experiments in the sealed modules serve as preliminary investigations for the advancement of future manned programs, including manned spacecraft and space stations. Furthermore, as part of a collaborative effort between China and Germany, the Shenzhou VIII manned spacecraft transported a closed ecosystem comprising ciliated chlorella and slender ruderal. The ciliated chlorella plays a vital role by generating oxygen for the astronauts on board. Simultaneously, the regular chlorella serves a dual function, not only providing essential oxygen but also serving as a nutritional food source, ensuring the astronauts receive complete and balanced nourishment while on the space station. Additionally, this pioneering ecosystem holds promising potential to offer valuable insights and solutions to numerous medical challenges faced in space exploration (http://www.wannengye.com/pages/QtwSlLIP).

The International Space Station had payloaded three exposure facilities, EXPOSE-E, EXPOSE-R, and EXPOSE-R2, to investigate the effect of the space environment (especially energetic radiations) on various chemical and biological samples. Experiments in the International Space Station involved studying the cyanobacterium *Chroococcidiopsis sp.* CCMEE 029's adaptation to Mars-like conditions, including ionizing radiation, atmosphere, and ultraviolet radiation. The results revealed that the dehydrated cyanobacterium successfully survived rehydration upon returning to Earth after a year and a half in a simulated Martian environment. This demonstrated that *Chroococcidiopsis sp.* CCMEE 029 possesses a degree of adaptability to the Martian environment(Napoli et al., 2022). Furthermore, the outcomes of acclimatization experiments with lichen (Circinaria gyrosa) conducted in environments emulating the Martian atmosphere, Martian soil analogs, and simulated Martian radiation demonstrate that lichen remains active in the Martian subsurface, where it is shielded from ultraviolet radiation (De La Torre Noetzel et al., 2020). In addition, the archaea experiments of methanogens have also been conducted on EXPOSE-R2 on the International Space Station, but the results have not yet been published (Rabbow et al., 2017).

In addition, the Advanced Closed Loop System (ACLS) is implemented on ISS. ACLS and NASA's Environmental Control and Life Support (ECLSS) program are the development of state-of-the-art technologies to guarantee astronauts' safety, comfort, and well-being throughout space missions. These initiatives emphasize the efficient utilization of resources and prioritize environmental sustainability to support the sustainability of space exploration endeavors.

Table 1. The experimental conditions, and simulations of other planets supported, and undertaken experiments in near-space, satellites and spacecraft, ISS and CSS. The ADAPT, LiFE and ROSE are the experiments related to the exo-ecosystem carried out aboard the ISS

(Bérces et al., 2015; Neuberger et al., 2015; Scalzi et al., 2012; Wassmann et al., 2012).

|  | Near-space | Satellites/Spacecraft | ISS | CSS |
| --- | --- | --- | --- | --- |
| Experimental condition | a. Low atmospheric pressure<br>b. Extremely low temperature<br>c. Humidity<br>d. A high flux of ultraviolet and cosmic radiation | a. Microgravity<br>b. Cosmic radiation | a. Microgravity<br>b. Cosmic radiation<br>c. Exposure platform | a. Microgravity<br>b. Cosmic radiation<br>c. Exposure platform with ultraviolet Shielding and Life-support temperature<br>d. In-orbit centrifuge |
| Simulations of other planets supported | a. Analogues of Mars' surface<br>b. Analogues of Venus' atmosphere |  | a. Analogues of Martian soil<br>b. Simulated Martian atmosphere<br>c. Simulated Martian radiation | d. Analogues of gravity of extraterrestrial planets<br>e. Analogues of Martian soil<br>f. Simulated Martian atmosphere<br>g. Simulated Martian radiation |
| Undertaken experiments | CAS-BAP<br>TC-BIOSEP |  | ADAPT<br>LiFE<br>ROSE | CHEESE(This Study) |

## 3. Design of the experiment of exo-ecosystem based on Chinese Space Station

The Exoplanet Habitability and Ecosystem Space Experiment (CHEESE) is an initiative planned for development on the Tiangong, China space station (CSS). The CHEESE encompasses three sections: ground-based experiments, Tiangong-based experiments, and transport processes (Fig. 1). The ground-based experiments section comprises matching tests, ergonomic experiments, synchronous comparison of spacial experiments and measurements and analysis of samples after return to the ground-based laboratory. The Tiangong-based experiments section contains the simulated gravity experiments and simulated radiation experiments. Following the six-month of the simulated gravity (Fig. 2 B1) and radiation experiments (Fig. 2 B2), the experimental samples will be preserved in refrigeration at 4°C (Fig. 2 B3) before their descent back to Earth. The upbound and downbound missions for transporting the spatial samples will be carried out by the Tianzhou and Shenzhou spacecraft (Fig. 2 C1\ C2), respectively.

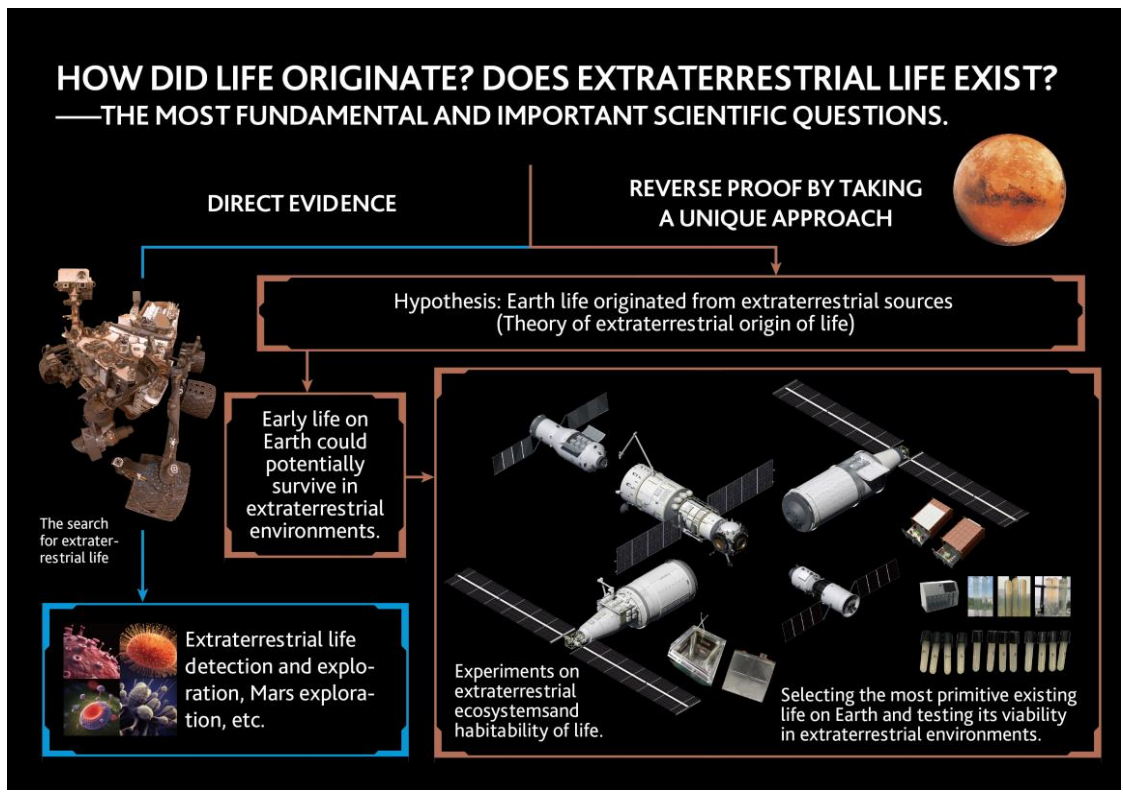

Figure 1. The conceptual design of China Exoplanet Habitability and Ecosystem Space Experiment (CHEESE) on China space station (CSS).

The matching tests conducted on ground-based laboratory include strain selection (Fig. 2 A1.1), anaerobic medium preparation (Fig. 2 A1.2), strain inoculation (Fig. 2 A1.3), Gas production measurement (Fig. 2 A1.4), survival time verification (Fig. 2 A1.5), temperature-matching experiments (Fig. 2 A1.6), and material compatibility testing (Fig. 2 A1.7), The strain selection is constrained by several factors, such as temperature range, metabolic mode, survival duration, and the presence of a genetic operating system. The anaerobic medium preparation is to select a more suitable medium state for space environment. We verified the adaptability of liquid medium, culture-free medium (strain slime), and solid medium, including rolling tube, slant, and full tube medium, to the spatial environment. Temperature-matching experiments were conducted to validate the gas production and viability of anaerobic archaea under varying temperature conditions. Pressure-matching test involved assessing the methane gas output by anaerobic archaaea and determining the range of pressure fluctuations that the anaerobic used in space experiments can endure. Material compatibility investigations focused on confirming the functional performance of anaerobic archaea when exposed to aerospace materials utilized in the space experiment. Given the potential duration of space experiments spanning up tubes to one year, the survival time of distinct cultures in the ground-simulated space environment was scrutinized to ensure their resilience and persistence throughout the experimental timeline.

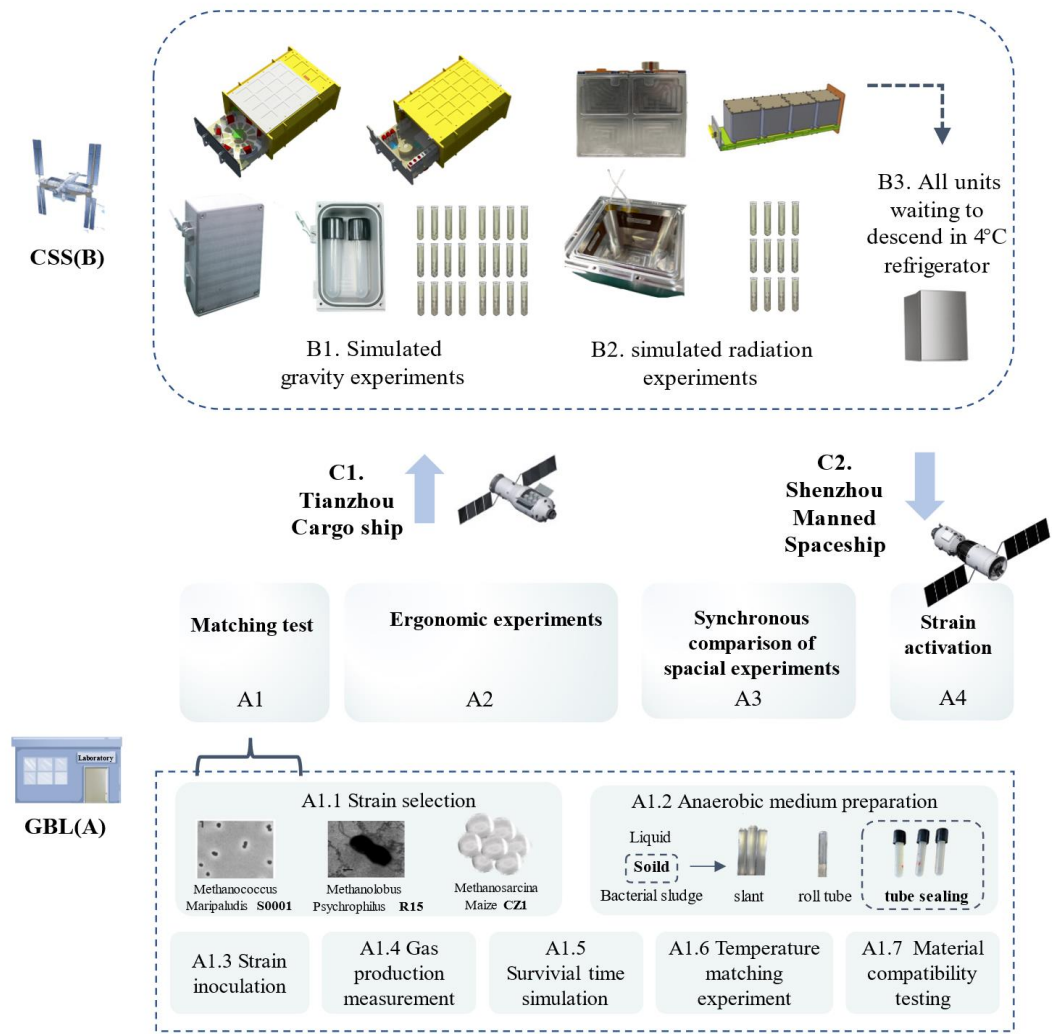

Figure 2. The roadmap of China Exoplanet Habitability and Ecosystem Space Experiment (CHEESE). CSS and GBL represent China's Space Station and Ground-based Laboratory, respectively.

The ergonomic experiments (Fig. 3) include the design and production of space experimental units, the design and production of the packages used by cargo ship, simulated mechanical experiments, and pilot operation training. The simulated gravity experimental unit is designed to meet the size of the mini-centrifuge, the sealing conditions of the anaerobic tubes, and the comfortable temperature for anaerobic archaea. The simulated radiation unit is designed to meet the size of the exposure platform, the requirements of the anaerobic archaea medium tubes to receive radiation, and the comfortable temperature for anaerobic archaea. The cargo package is to protect the experimental unit during the upward and downward transport processes. The mechanics experiments are to simulate the impact conditions that may be encountered during the upward and downward travels.

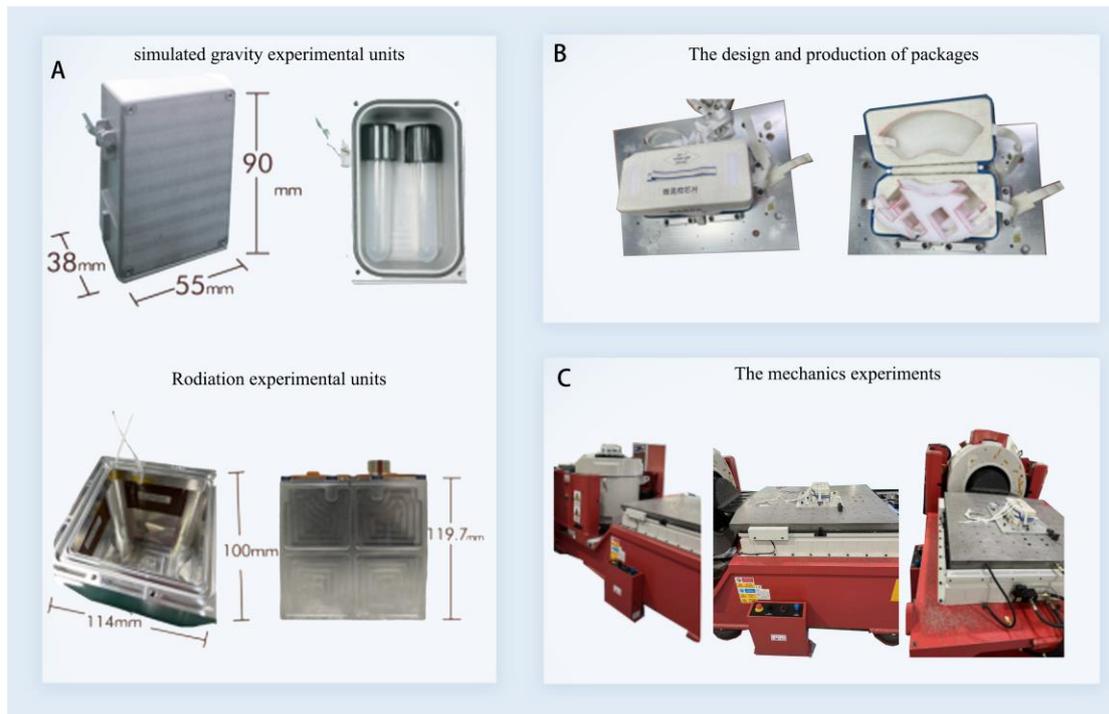

Figure 3. The design and production of space experimental units (A), the design and production of the packages (B) and the simulated mechanical experiments.

The simulated gravity experiment will be conducted on Tiangong using a compact centrifuge located within the life and ecology cabinet in the Wentian module. The mini-centrifuge equipment encompasses both centrifuge module and static module (Fig. 4) and maintains a controlled temperature range of 19-35 °C. The centrifuge module is capable of simulating gravity levels comparable to Mars or the Moon (0.4g/ 0.2g), while the static module can create a microgravity environment representative of space conditions ( ~ 0.01g), operating at the same temperature as the centrifuge module. The simulated radiation experiments on the space station will be conducted on an extravehicular exposure platform in the Mengtian module. The extravehicular exposure platform provides a life support temperature of 5 to 20 °C and a selectable UV radiation environment. In order to monitor the radiation received by the spacial samples, we have installed CR-39 in the radiation experimental units and TLDs on medium tubes in the units.

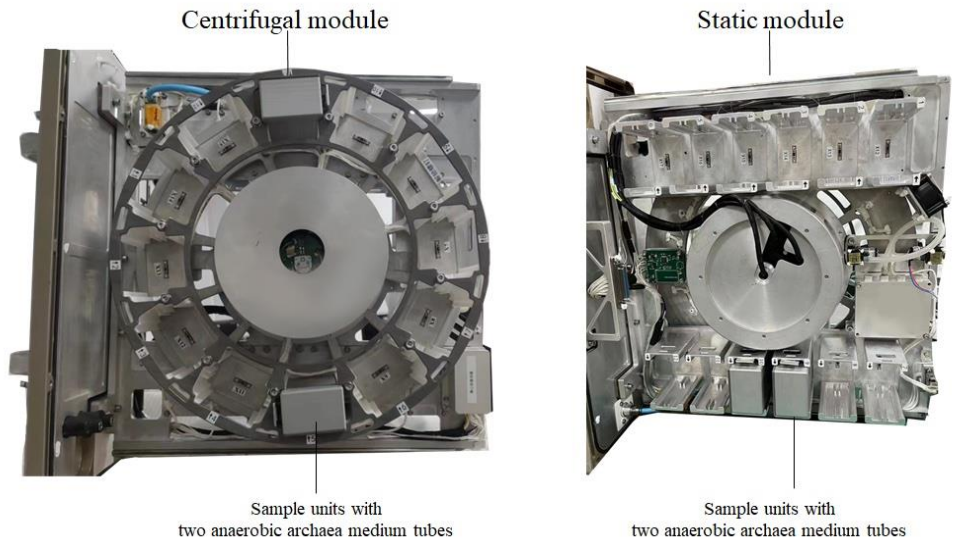

Figure 4. The mini-centrifuge equipment with centrifugal module and static module on ISS.

## 4. Future Prospects and Conclusions

Exploring the possibility of constructing exo-ecosystems is an exciting frontier in space exploration. While significant progress has been made in our understanding of space and the potential for life beyond Earth, there is still much to discover. The construction of exo-ecosystems opens up new avenues for scientific exploration, resource utilization, and preservation of Earth's biodiversity.

In the coming years, it is crucial to identify potential targets for exo-ecosystem construction. Bodies such as the Moon, Mars, and other celestial bodies with the potential to support life or sustain human habitation should be prioritized for further research and development. Additionally, interdisciplinary collaboration among scientists, engineers, biologists, and environmentalists is essential to advance our knowledge and capabilities in this field.

To realize the full potential of exo-ecosystems, it is recommended that future research and development focus on several key areas. These include advancing closed-loop life support systems, optimizing resource utilization technologies, studying the long-term effects of living in extraterrestrial environments, and developing sustainable practices for exo-ecosystem operation.

In conclusion, exploring the possibility of constructing exo-ecosystems is a crucial step in our quest for space exploration and colonization. The scientific discoveries, resource utilization potential, preservation of Earth's biodiversity, socioeconomic benefits, and technological advancements offered by exo-ecosystems are compelling reasons to pursue this endeavor. As we venture beyond Earth, it is imperative to invest in research, development, and ethical considerations to ensure the sustainability and viability of exo-ecosystems. By doing so, we not only expand our understanding of the universe but also pave the way for a future where humanity can thrive beyond the confines of our home planet.

**Reference**

Bérces, A., Egyeki, M., Fekete, A., Horneck, G., Kovács, G., Panitz, C., Rontó, Gy., 2015. The PUR Experiment on the EXPOSE-R facility: biological dosimetry of solar extraterrestrial UV radiation. Int. J. Astrobiol. 14, 47–53. https://doi.org/10.1017/S1473550414000287

Caporale, A.G., Palladino, M., De Pascale, S., Duri, L.G., Rouphael, Y., Adamo, P., 2023. How to make the Lunar and Martian soils suitable for food production - Assessing the changes after manure addition and implications for plant growth. J. Environ. Manage. 325, 116455. https://doi.org/10.1016/j.jenvman.2022.116455

Dachev, T.P., Bankov, N.G., Tomov, B.T., Matviichuk, Yu.N., Dimitrov, Pl.G., Häder, D.-P., Horneck, G., 2017. Overview of the ISS Radiation Environment Observed during the ESA EXPOSE-R2 Mission in 2014-2016: ISS Radiation Environment. Space Weather 15, 1475–1489. https://doi.org/10.1002/2016SW001580

De La Torre Noetzel, R., Ortega García, M.V., Miller, A.Z., Bassy, O., Granja, C., Cubero, B., Jordão, L., Martínez Frías, J., Rabbow, E., Backhaus, T., Ott, S., García Sancho, L., De Vera, J.-P.P., 2020. Lichen Vitality After a Space Flight on Board the EXPOSE-R2 Facility Outside the International Space Station: Results of the Biology and Mars Experiment. Astrobiology 20, 583–600. https://doi.org/10.1089/ast.2018.1959

Kasiviswanathan, P., Swanner, E.D., Halverson, L.J., Vijayapalani, P., 2022. Farming on Mars: Treatment of basaltic regolith soil and briny water simulants sustains plant growth. PLOS ONE 17, e0272209. https://doi.org/10.1371/journal.pone.0272209

Li, Y., Sun, Y., Zheng, H., Shang, P., Qu, L., Lei, X., Liu, H., Liu, M., He, R., Long, M., Sun, X., Wang, J., Zhou, G., Sun, L., 2021. Recent Review and Prospect of Space Life Science in China for 40 Years. Chin. J. Space Sci. 41, 46. https://doi.org/10.11728/cjss2021.01.046

Lin, W., He, F., Zhang, W., Yao, Z., Shen, J., Ren, Z., Yuan, H., Cai, R., Wei, Y., Pan, Y., 2022. Astrobiology at altitude in Earth's near space. Nat. Astron. 6, 289–289. https://doi.org/10.1038/s41550-022-01606-1

Maus, D., Heinz, J., Schirmack, J., Airo, A., Kounaves, S.P., Wagner, D., Schulze-Makuch, D., 2020. Methanogenic Archaea Can Produce Methane in Deliquescence-Driven Mars Analog Environments. Sci. Rep. 10, 6. https://doi.org/10.1038/s41598-019-56267-4

Mickol, R., Laird, S., Kral, T., 2018. Non-Psychrophilic Methanogens Capable of Growth Following Long-Term Extreme Temperature Changes, with Application to Mars. Microorganisms 6, 34. https://doi.org/10.3390/microorganisms6020034

Mickol, R.L., Kral, T.A., 2017. Low Pressure Tolerance by Methanogens in an Aqueous Environment: Implications for Subsurface Life on Mars. Orig. Life Evol. Biospheres 47, 511–532. https://doi.org/10.1007/s11084-016-9519-9

Napoli, A., Micheletti, D., Pindo, M., Larger, S., Cestaro, A., De Vera, J.-P., Billi, D., 2022. Absence of increased genomic variants in the cyanobacterium Chroococcidiopsis exposed to Mars-like conditions outside the space station. Sci. Rep. 12, 8437. https://doi.org/10.1038/s41598-022-12631-5

Neuberger, K., Lux-Endrich, A., Panitz, C., Horneck, G., 2015. Survival of Spores of *Trichoderma*

*longibrachiatum* in Space: data from the Space Experiment SPORES on EXPOSE-R. Int. J. Astrobiol. 14, 129–135. https://doi.org/10.1017/S1473550414000408

Rabbow, E., Rettberg, P., Barczyk, S., Bohmeier, M., Parpart, A., Panitz, C., Horneck, G., Burfeindt, J., Molter, F., Jaramillo, E., Pereira, C., Weiß, P., Willnecker, R., Demets, R., Dettmann, J., Reitz, G., 2015. The astrobiological mission EXPOSE-R on board of the International Space Station. Int. J. Astrobiol. 14, 3–16. https://doi.org/10.1017/S1473550414000202

Rabbow, E., Rettberg, P., Barczyk, S., Bohmeier, M., Parpart, A., Panitz, C., Horneck, G., Von Heise-Rotenburg, R., Hoppenbrouwers, T., Willnecker, R., Baglioni, P., Demets, R., Dettmann, J., Reitz, G., 2012. EXPOSE-E: An ESA Astrobiology Mission 1.5 Years in Space. Astrobiology 12, 374–386. https://doi.org/10.1089/ast.2011.0760

Rabbow, E., Rettberg, P., Parpart, A., Panitz, C., Schulte, W., Molter, F., Jaramillo, E., Demets, R., Weiß, P., Willnecker, R., 2017. EXPOSE-R2: The Astrobiological ESA Mission on Board of the International Space Station. Front. Microbiol. 8, 1533. https://doi.org/10.3389/fmicb.2017.01533

Scalzi, G., Selbmann, L., Zucconi, L., Rabbow, E., Horneck, G., Albertano, P., Onofri, S., 2012. LIFE Experiment: Isolation of Cryptoendolithic Organisms from Antarctic Colonized Sandstone Exposed to Space and Simulated Mars Conditions on the International Space Station. Orig. Life Evol. Biospheres 42, 253–262. https://doi.org/10.1007/s11084-012-9282-5

Sinha, N., Nepal, S., Kral, T., Kumar, P., 2017. Survivability and growth kinetics of methanogenic archaea at various pHs and pressures: Implications for deep subsurface life on Mars. Planet. Space Sci. 136, 15–24. https://doi.org/10.1016/j.pss.2016.11.012

Smith, D.J., 2013. Microbes in the Upper Atmosphere and Unique Opportunities for Astrobiology Research. Astrobiology 13, 981–990. https://doi.org/10.1089/ast.2013.1074

Wamelink, G.W.W., Frissel, J.Y., Krijnen, W.H.J., Verwoert, M.R., Goedhart, P.W., 2014. Can Plants Grow on Mars and the Moon: A Growth Experiment on Mars and Moon Soil Simulants. PLoS ONE 9, e103138. https://doi.org/10.1371/journal.pone.0103138

Wang, Y., Jiang, Y., Sun, Z., Wang, C., 2023. The Temperature-Controlled Biological Samples Exposure Payload(TC-BIOSEP) for Balloon-Based Astrobiology Research. Microgravity Sci. Technol. 35, 10. https://doi.org/10.1007/s12217-023-10035-2

Wassmann, M., Moeller, R., Rabbow, E., Panitz, C., Horneck, G., Reitz, G., Douki, T., Cadet, J., Stan-Lotter, H., Cockell, C.S., Rettberg, P., 2012. Survival of Spores of the UV-Resistant *Bacillus subtilis* Strain MW01 After Exposure to Low-Earth Orbit and Simulated Martian Conditions: Data from the Space Experiment ADAPT on EXPOSE-E. Astrobiology 12, 498–507. https://doi.org/10.1089/ast.2011.0772